# Perylene-Based Non-Covalent Functionalization of 2D Materials

*Mario Marcia, Andreas Hirsch, Frank Hauke**

Department of Chemistry and Pharmacy and Institute of Advanced Materials and Processes (ZMP), Friedrich-Alexander University of Erlangen-Nuremberg
Henkestrasse 42, 91054 Erlangen, Germany
Fax: +49 (0) 911 650 78 650 15
E-mail: frank.hauke@fau.de

## 1. Abstract

The surfactant assisted exfoliation and non-covalent functionalization of two-dimensional layered materials, like graphene, transition metal dichalcogenides, etc., has extraordinarily been propelled forward within the last 5 years. Numerous molecules have been designed and attached to the exfoliated layers, and based on their outstanding properties, perylene based dyes have become one of the most frequently used π-detergents. Therefore, the prospect of this micro review is to summarize the most prominent achievements in this rapidly progressing field of research.

## 2. Introduction

Graphene,[1] the prototype of novel 2D materials, has become a highly promising icandidate for future carbon-based devices.[2] Nowadays, the most important chemical routes towards its production are the liquid-phase exfoliation of graphite,[3-4] the reduction of graphene oxide,[5-6] and its bottom-up synthesis, starting from small precursor molecules.[7]

Besides its production, also its functionalization has attracted a lot of interest in the scientific community due to its unprecedented physical properties and the wide range of possible applications where this carbon nano-material and its derivatives can be implemented. With this objective, the covalent functionalization of graphene has been investigated in detail[8] and also the non-covalent modification of graphene and other 2D materials has extraordinarily been propelled forward within the last 5 years.

With respect to this non-covalent functionalization approach, numerous molecules have been designed and attached to the extended π-honeycomb lattice of graphene and graphene oxide (GO). As recently reviewed by Ciesielski *et al*.[3] and by Georgakilas *et al*.,[9] among the vastness of functionalizing reagents proposed, polycyclic aromatic hydrocarbons (PAHs) with their characteristic structural motif have been identified as most promising surfactant/functionalization reagents. In particular, pyrene- and naphthalene-based derivatives and perylene-based molecules have been developed as one of the most frequently used π-detergents.

Although the physics and chemistry of perylene-based dyes have thoroughly been analyzed in the last ten years and many molecules belonging to this class of compounds have been used for the exfoliation and non-covalent functionalization of graphene, surprisingly, this topic has not been reviewed yet. Therefore, the aim of this micro review is to summarize the most prominent achievements in this field of research.

## 2. Discussion

### 2.1 The Early Work

Due to their outstanding physical and optical properties and due to their propensity towards self-assembly, perylene-based dyes have been implemented in many fields of technology, as reported by Würthner *et al*.[10] The rigid aromatic perylene unit (Figure 1) is a very suitable scaffold for a subsequent chemical derivatization, both at its axial and equatorial positions. For example, upon attachment of two dicarboxylic acid groups, the perylene core can be axially extended to build up the structure of perylene-3,4,9,10-tetracarboxylic dianhydride (**PTCDA**). By additional functionalization with primary amines, the generic parent perylene diimide (**PDI**) chromophore can be obtained.[11] Moreover, the chemical modification can also take place in the equatorial region, both at the ortho[12] and bay[13] positions.

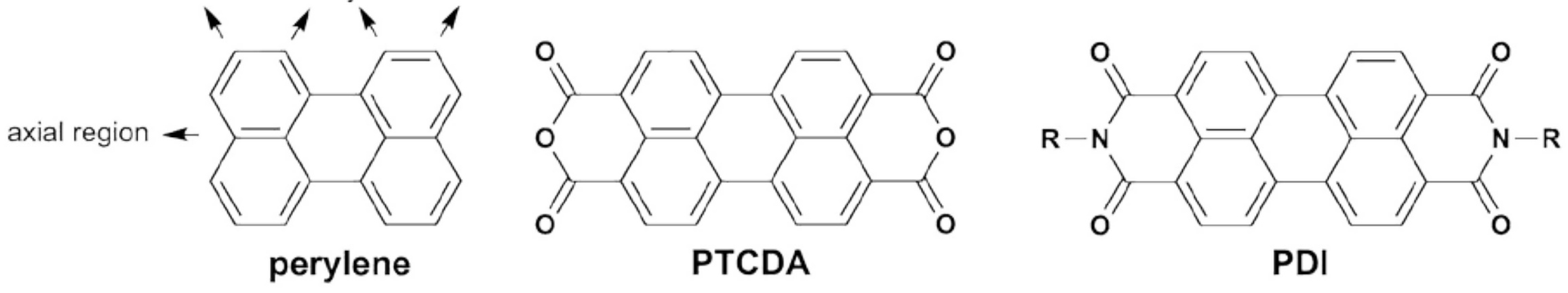

**Figure 1** Chemical structure of **perylene**, **PTCDA**, and a generic **PDI**.

Since the early 1990s the interaction of perylene-based derivatives has been studied on graphite by means of scanning tunneling microscopy (STM).[14] With the discovery of graphene in 2004[1] the interaction of such derivatives with the novel synthetic 2D carbon allotrope has also been pursued. Independently, the group of Ley[15] and of Hersam[16] reported on the assembly and the structure of PTCDA deposited on bilayer graphene at 4.7 K and on a single layer of graphene (SLG) at room temperature, respectively.

Concomitantly with the availability of these works, the first articles concerning the use of perylene-based molecules for the wet-chemical exfoliation and non-covalent functionalization of graphene, GO, and reduced graphene oxide (rGO) were published.

For example, Li *et al*.[17] employed the hydrolyzed form of **PTCDA** - 3,4,9,10-perylene tetracarboxylic acid **(PTCA) -** for the functionalization of rGO (Figure 2). The four carboxylic acid groups of **PTCA** served as anchor points for the subsequent *in situ* growth of ionic-liquid stabilized and positively charged gold nanoparticles (NP). Their successful decoration was confirmed by X-ray photoelectron spectroscopy (XPS) and X-Ray diffraction (XRD) measurements, while analysis by transmission electron microscopy (TEM) suggested a homogeneous coverage of the Au-NP on the flakes of **PTCA**-functionalized rGO. Additionally, the electrocatalytic behavior towards oxygen reduction of the rGO/**PTCA**/Au-NP hybrid was tested. As indicated by cyclic voltammetry (CV) measurements, highly promising applications of this hybrid nano-structure could be envisaged in the field of chemical sensors.

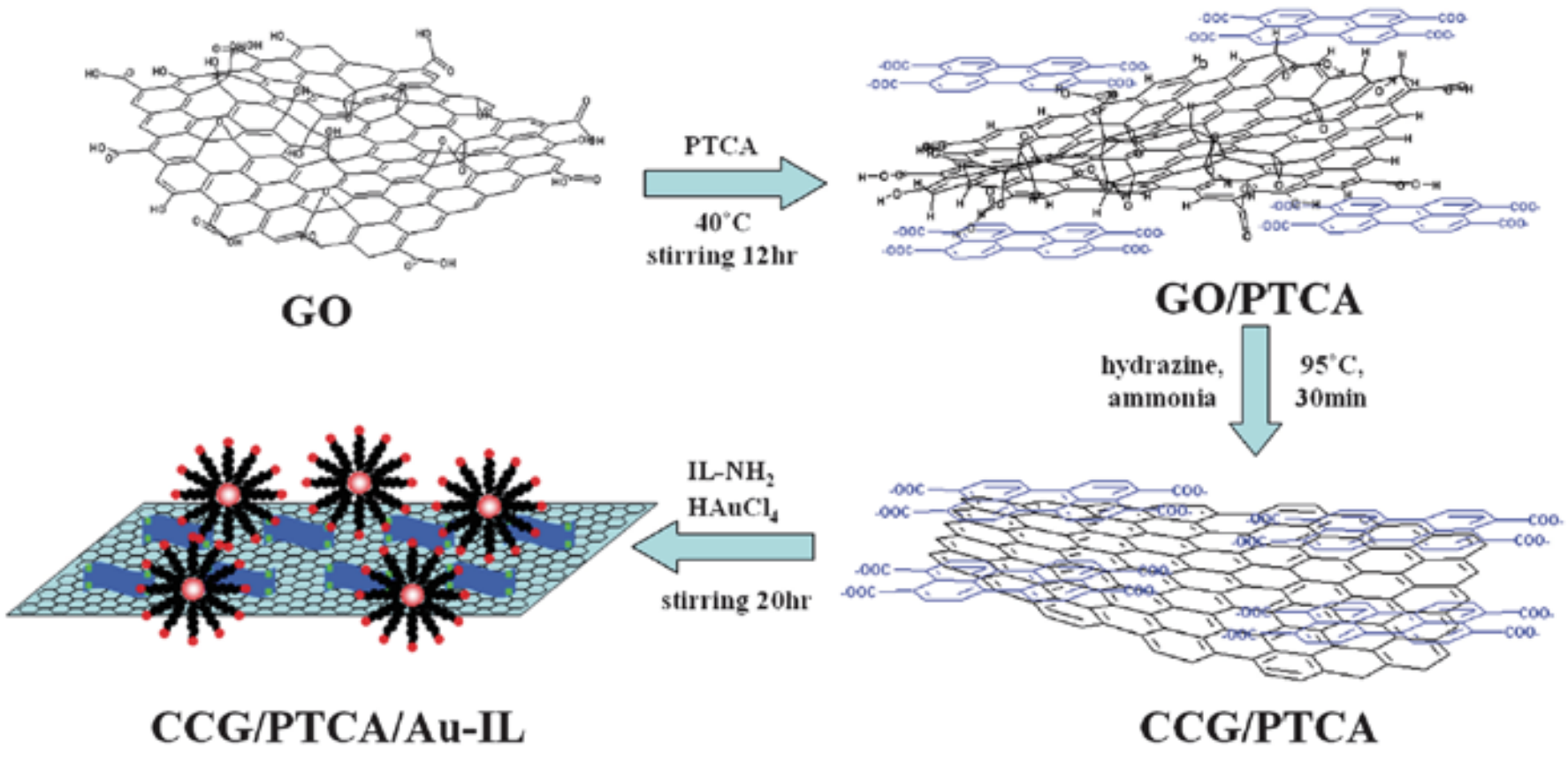

**Figure 2.** Illustration of the preparation of chemically converted graphene (**CCG**) decorated with **PTCA** and gold/ionic liquid nanoparticles (**AU-IL**). Ref.[17] with permission from Royal Society of Chemistry, Copyright 2009.

At the same time, Su *et al*.[18] reported the use of a pyrene- and a PDI-based anionic surfactant for the stabilization of aqueous dispersions of rGO nano-sheets. Careful analysis by atomic force microscopy (AFM) provided the proof for a strong non-covalent functionalization and a sandwich model structure was proposed to explain the attachment of the $\pi$-surfactants on both sides of the rGO sheets. Furthermore, as shown both by Raman and XPS measurements, the different chemical nature of the surfactants (pyrene as donor; PDI as acceptor) allowed a precise tuning of the electronic properties of the functionalized graphene flakes.

Also in 2009, Englert *et al*.[19] reported the use of a 2$^{nd}$ generation Newkome dendronized PDI-based bolaamphiphilic molecule for the stabilization and dispersion of pristine exfoliated turbostratic graphite in aqueous solution. Even though no evidence of strong interaction between the PDI and the delaminated carbon nano-material could be detected in solution, the presence of single layer graphene among the products of the exfoliation could be proven by means of Raman spectroscopy and AFM measurements.

Further pioneering work concerning the non-covalent functionalization of graphene was published one year later by Kozhemyakina *et al*.[20] This time an asymmetric neutral PDI-based dye (**an-PDI**) was used to aid the exfoliation of turbostratic graphite in *N*-methyl-2-pyrrolidone (NMP) by gentle magnetic stirring (Figure 3).

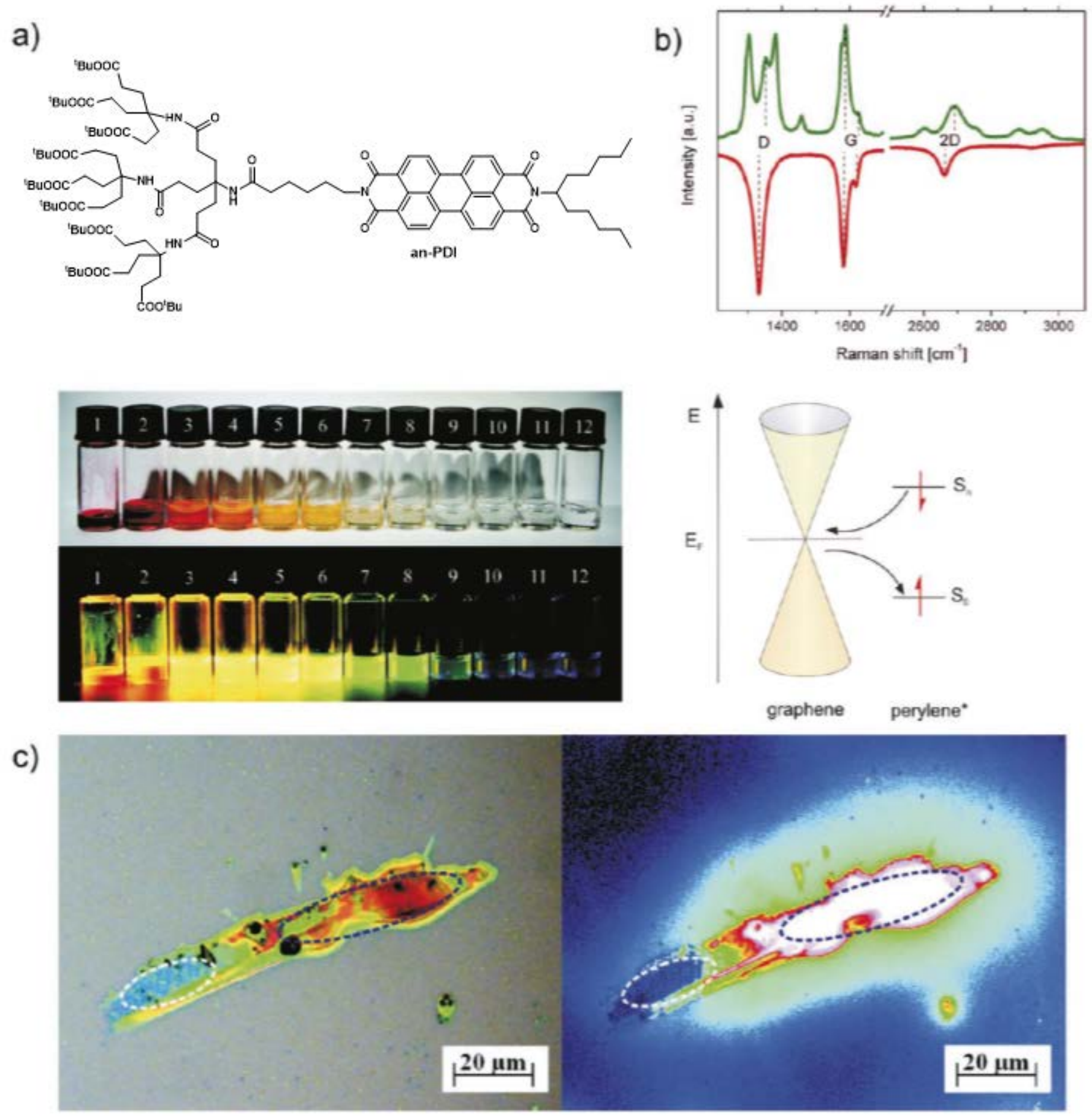

**Figure 3.** a) Bolaamphiphilic perylene bisimide **an-PDI** and its solutions in NMP in the concentration range from $10^{-2}$ M to $10^{-7}$ M at day light and at 366 nm excitation. b) Top: Raman spectra of the dispersion of turbostratic graphite with **1** in NMP. At 532 nm excitation (green spectrum), both characteristic single- and multi-layer graphene peaks and peaks originating from **an-PDI** are observed. Bottom: Schematic representation of the graphene band structure in the vicinity of the K-point in the 1st B.Z. with empty states above and filled states below the Fermi Energy (EF) together with the photoexcited singlet state of the perylene chromophore. c) Micrographs of graphene-perylene dispersion deposited on a Si/$SiO_2$ 300 nm coated substrate. Left: White light illumination. Blue contrast corresponds to graphene and finely dispersed graphite particles (marked with white oval), yellow-orange contrast – to perylene **an-PDI** (marked with blue oval). Right: Fluorescence (green light excitation at 545 nm, emission from 605 nm) is observed for unbound perylene (white, red, yellow and light blue color code, marked with blue oval) and is quenched for **an-PDI** interacting with graphite and graphene (marked with a white oval). Adapted form ref.[20] with permission from Wiley-VCH Verlag GmbH & Co, Copyright 2010.

After centrifugation, a homogeneous and long-term stable dispersion of few- and single layer graphene could be obtained. Upon deposition on Si/$SiO_2$ wafers, the interaction of the PDI dye with the graphene deposits could be detected by Raman spectroscopy (Figure 3b). Moreover, the effects of the non-covalent functionalization could be visualized by means of fluorescence microscopy in the solid state as well as by fluorescence spectroscopy in solution (Figure 3c). As a matter of fact, careful titration experiments showed that the quantum yield of the **an-PDI** could be reduced of about 35% by addition of dispersed graphitic nano-material.

Additionally, the group of Loh[21-22] employed a commercially available neutral PDI dye (**PDI-C$_8$**) to prepare **PDI-C$_8$**/rGO wires by a hydrothermal treatment of GO in *N*,*N*-

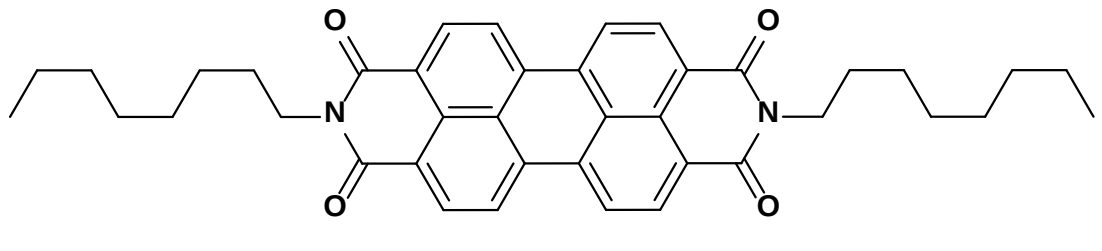

**PDI-C$_8$**

dimethylformamide (DMF). The as-prepared wires proved to exhibit a cylindrical shape, being several hundreds of micrometer long, with a diameter comprised in the range of 5-10 µm and to be highly crystalline. Moreover, these systems are characterized by a strong anisotropy and their photoluminescence resulted to be strongly polarization-dependent. Furthermore, the presence of the rGO as outer coating layer of the wires could be confirmed by means of fluorescence microscopy and Raman spectroscopy. Finally, the photovoltaic properties of the **PDI-C$_8$**/rGO wires were investigated. Due to the more efficient charge transport within the assembly, the **PDI-C$_8$**/rGO wires were found to exhibit far superior performances, such as better current density values as well as an improved power conversion efficiency, in comparison to **PDI-C$_8$** alone or discrete mixtures of **PDI-C$_8$** and rGO.

The experimental results presented in these early articles matched the previous observations concerning the individualization and functionalization of single wall carbon nanotubes (SWCNTs) with PDI-based surfactants.[23] The physisorption of perylene-based molecules onto both carbon allotropes in solution resulted usually in the presence of strong interactions due to $\pi$–$\pi$ stacking interactions, which translated in a large bathochromic shift of the absorption and emission features of the dyes. Additionally, the fluorescence of the chromophore was dramatically quenched as a result of energy and photo-induced electron transfers (PETs).[24-26] As a matter of fact, although non-disruptive, the non-covalent functionalization determines a significant change in the density of states of the graphene substrate. The modulation of its electrical properties usually results in strong doping effects: *p*-doping for PTCA and PDI derivatives, while *n*-doping for perylene bearing electron donating groups.[27] On the basis of calculations by means of density functional theory (DFT), this modification could be further explained by the fact that the stacking of perylene-based derivatives on graphene can open and tailor its band gap.[28-29]

Since 2010, approximately one-hundred reports (Scifinder®) were published dealing with the interaction of perylene-based derivatives with graphene. Here, several reports focused on the physical aspects of the perylene-based devices and the perylene/graphene interaction.[30-43] Moreover, some derivatives bearing reminiscent perylene structures have also been suggested for the modification of graphene.[44-45]

However, in the following we will focus on the discussion of the pure wet-chemical approaches which led to the successful implementation of classic perylene- and PDI-based derivatives in graphene-containing materials. In order to better review the state-of-the-art of this technology, the literature has been compared according to the type of molecule employed as functionalizing agent, independently on the kind of 2D material which was functionalized and on the year of publication.

### 2.2 PTCDA and PTCA as Surfactant Molecules

Since **PTCDA** might be defined as the starting building block for the synthesis of further perylene- and PDI-based derivatives, its role as functionalizing agent will be described at first. In the work of Peng *et al*.[46] **PTCDA** was used to modify graphene deposited on the surface of a polished gold electrode to develop a sandwich-type electrochemical aptasensor for the detection of thrombin. The successful fabrication of the biosensor could be followed stepwise by means of CV measurements and electrochemical impedance spectroscopy. By comparing the peak current for the bare and the functionalized electrode, the authors determined that the modification with the **PTCDA**/graphene composite could lead to an increase of the electroactive surface area and to a superior conductivity. Moreover, Zhu *et al*.[47] employed small aromatic molecules such as **PTCDA**, **PTCDI** (obtained by oxidative coupling of two naphthanlene-1,8-dicarboxylic acid imides[48]), and 1-aminopyrene (**1-AP**) for the non-covalent functionalization of graphite nanoplatelets in ethylene glycol/water mixtures. These moieties were also used as anchoring points for the *in situ* deposition of Pt-NP. From the point of view of TEM and XRD characterizations, **1-AP** and **PTCDA** exhibited better results concerning the size and distribution of the Pt-NP on the surface of the exfoliated nanoplatelets, compared to **PTCDI**. The authors attributed this behavior to the different chemical structures of the three aromatic derivatives. The composite materials were also tested as anode catalysts for applications in direct methanol fuel cells. The results of the CV and chronoamperometry measurements showed that the catalytic activity and the stability were maximized when **1-AP** or **PTCDA** were used, in perfect agreement with the observations gathered from TEM and XRD. Recently, Cui *et al*.[49] proposed a composite of rGO wrapped with **PTCDA** as high-performance organic cathode for lithium batteries. Here, **PTCDA** was dispersed in water in presence of GO and stirred at 40 °C to obtain a homogeneous slurry. After drying, the composite was thermally reduced (200 °C < T < 500 °C). Scanning electron microscopy (SEM) and TEM measurements revealed that **PTCDA** distributes on GO as submicron rods. After reduction to rGO, the small aromatic molecule could be more densely incorporated and a laminated structure as a highly conductive 3D network could be formed (Figure 4). Furthermore, the wrapped composite structure could reduce the contact between the bare **PTCDA** and the electrolyte, thus suppressing side reactions at the electrode/electrolyte interface. The electrochemical performances of the **PTCDA**/rGO composite were found to be superior to similar **PTCDA**-based organic cathodes, with a coulombic efficiency above 98% and a quite high Li-ion diffusion rate.

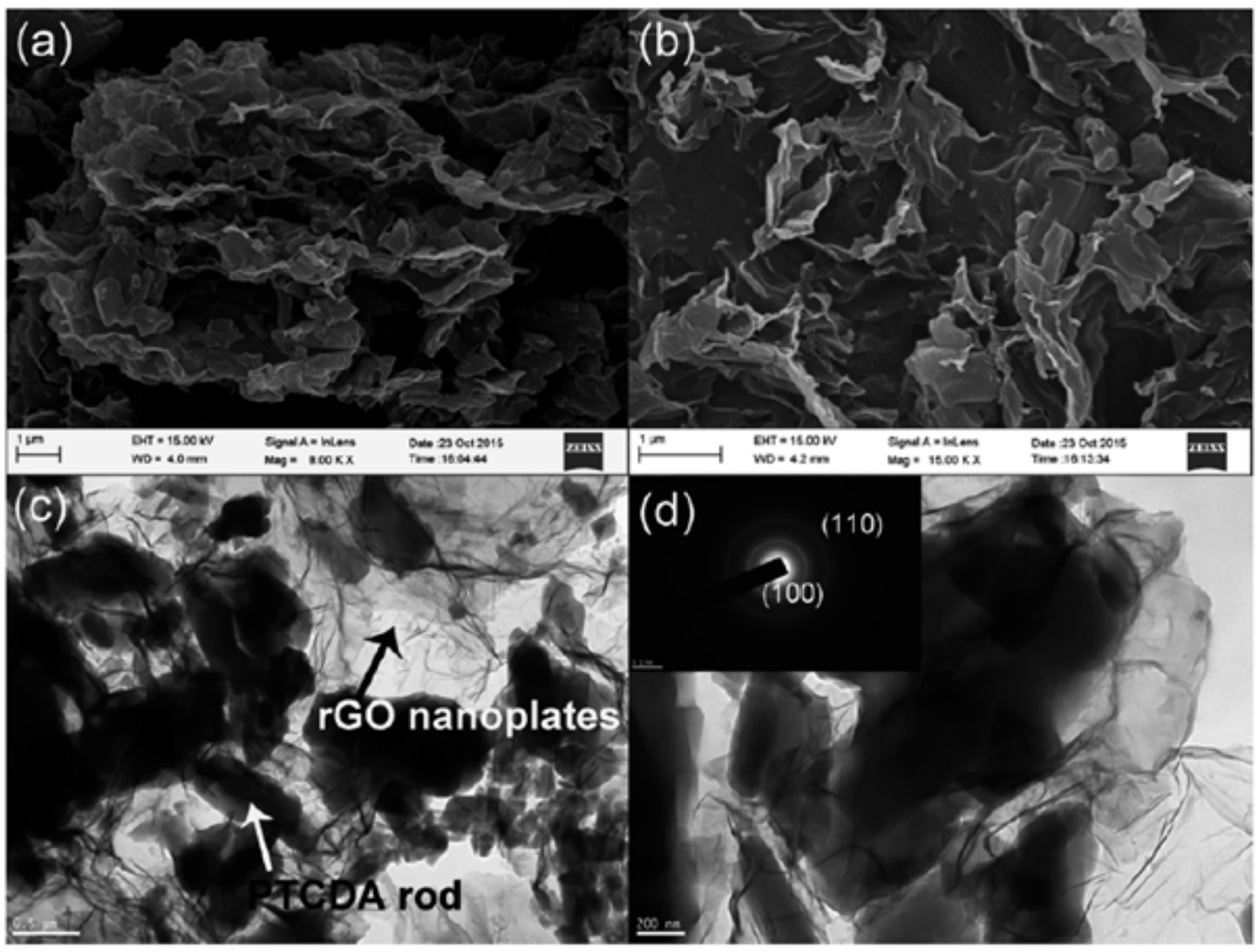

**Figure 4.** a and b) SEM images of the PTCDA/rGO composite prepared at 400 °C in different magnifications. c and d) TEM image of the rGO wrapped PTCDA composite. Inset shows the corresponding SAED pattern. Ref.[49] with permission The Royal Society of Chemistry, Copyright 2016.

Due to its simple aromatic structure and straightforward synthesis, **PTCA** has been most commonly used for the preparation of perylene-functionalized graphene composite materials. In detail, it has found a broad application as key building block in the fabrication of label-free electrochemical aptasensors for the detection of cancer cells[50] or thrombin.[51] Moreover, it has been implemented a) in electrochemilumines-cent aptasensors for the detection of thrombin,[52-53] b) in label-free electrochemical biosensors for DNA analysis,[54] c) in electrochemical sensors for the detection of ascorbic acid, dopamine, uric acid, and tryptophan,[55-56] d) in photoelectrochemical aptasensors for the detection of mercury(II) ions,[57-58] e) in electrochemical biosensor for the detection of hydrogen peroxide,[59] f) in electrochemical immunosensor for the detection of 3,3‘,5-triiodothyronine, a diagnostic marker of thyroid disease,[60] and 5-hydroxymethylcytosine, a cancer biomarker[61] as well as in g) a ruthenium(II)-based electrochemiluminescent immunosensor for the detection of α-fetoprotein, a cancer biomarker.[62]

Beside these sensing applications, due to its high biocompatibility **PTCA** has been employed to stabilize rGO in order to design novel nano-materials for biomedical applications. As explained by Gan and co-workers,[63] the interaction of **PTCA**/rGO composites with mainly four types of blood proteins (human fibrinogen, γ-globulin, bovine serum albumin, and insulin) could be investigated. In detail, the hydrophobic amino acids of the proteins were found to act as fluorescence markers to provide information about the binding order with the graphene-based platform. Further information concerning the binding order could be gathered by means of AFM as well as circular dichroism. Here, the size, shape, and type of driving forces also play a crucial role in the binding ability. Altogether it was found that the interaction of the **PTCA**/rGO hybrids with the human blood proteins depended strongly on the number and density of the surface aromatic amino acids of the protein as well as to the presence of electrostatic repulsions.

Moreover, **PTCA**/graphene substrates were synthesized for applications in direct ethanol fuel cells and biofuel cells. Li *et al*.[64] proposed a catalyst for electro-oxidation of ethanol based, on a **PTCA**/graphene carrier decorated with Pd-NP. As already reported in the work of Zhu *et al.*,[47] the presence of small organic molecules bearing amino-, oxo-, and hydroxy functionalities can help to improve the distribution of the NP on top of graphene layers. Moreover, in the experiments of Li and coworkers three different reducing agents were compared and the best results concerning the size, distribution, and crystallinity of the Pd-NP were obtained with sodium borohydride. The results of the structural analysis (XRD, TEM) were also confirmed by CV and chronoamperometric measurements, which stated that by employing sodium borohydride as reducing agent the composite with the best electrocatalytic activity, the largest electrochemically active surface area, and fastest kinetics could be achieved. Zhang *et al*.[65], instead, synthesized a 3D-graphene network with a Ni(II)-exchange/KOH activation method using a sulfonic acid ion-exchange resin as carbon precursor and further functionalized it by ultra-sonication in presence of **PTCA**. Due to its larger specific surface area, as a result of its open and interconnected structure, this special kind of 3D-graphene proved to be a better substrate in comparison to classically dispersed and functionalized graphene sheets. By a successive DCC-coupling with dopamine, the composite material 3D-graphene/**PTCA**/dopamine could be obtained and deposited on a glassy carbon electrode to work as the biocatode in a fuel cell.

Furthermore, wet chemical approaches were used to investigate the interaction between **PTCA** and graphene. On the one hand, a tailor-made heat-assisted procedure was developed by Gan *et al*.[66] in order to promote the formation of different supramolecular nano-structures between **PTCA** and GO or rGO. Under the selected experimental conditions of their work, GO was reported to induce J-stacking of the dye while rGO H-stacking. The difference in the self-assembly of the dye could be well evidenced from AFM and TEM measurements, where the formation of tubular structures for **PTCA** on the surface of GO and small-size nano-buds for **PTCA** on the surface of rGO could be observed. Additionally, taking advantage of the photo-activity of **PTCA**, two types of photocurrent configurations in both solid and liquid phase could be successfully tested, thus highlighting the potential of **PTCA**/graphene hybrids for applications in the field of optoelectronics and as a component in solar cells. On the other hand, Li *et al*.[67] employed the sodium salt of **PTCA** as stabilizer in a novel process to improve the dispersion yield of rGO in water. The authors compared two batches of GO and treated one of them with sodium hydroxide to remove highly oxidized carbonaceous adsorbates from the surface of the oxygen-functionalized graphene sheets. After reduction to rGO, **PTCA** was added to both batches in order to help the stabilization of the dispersed reduced nano-sheets. Indeed a concentration as high as 1 mg/mL could be reached for the batch treated in alkaline conditions while for the untreated sample the highest concentration in the dispersion was 0.2 mg/mL, thus justifying the postfunctional treatment of GO to enhance the dispersibility of rGO under aqueous conditions.

Recently, Wang *et al*.[68] and Gan *et al*.[69] proposed the use of CV measurements to study the electrochemically driven protonation/deprotonation equilibria of **PTCA** immobilized on the surface of rGO. Four voltammetric peaks appear between -0.1 V and -0.6 V (*vs.* saturated calomel electrode) or between -0.05 V and -0.7 V (*vs.* silver chloride electrode) which can be correlated to the dissociation

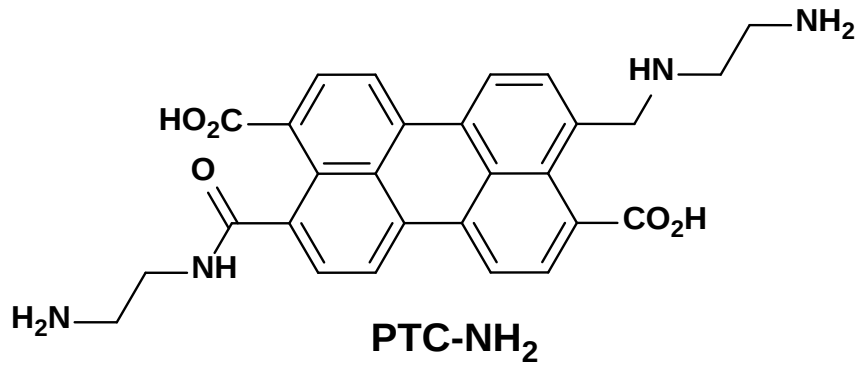

**PTC-NH$_2$**

of the four protons of **PTCA**. The reason for the presence of these unique electrochemical processes is quite complex and will not be discussed here. However, due the ultrafast charge-discharge ability of the **PTCA**/rGO hybrid its feasible application in a supercapacitor prototype was lately presented, which showed a high gravimetric capacitance and a power density higher than lithium-ion batteries.

In addition to **PTCA**, another perylene-based derivative which has been extensively employed for the functionalization of graphene is **PTC-NH$_2$**. This molecule can be synthesized by reaction of **PTCDA** with ethylenediamine. As in the case of **PTCA**, the most common application for **PTC-NH$_2$**/graphene conjugates is found in the field of sensors. Its use in electrochemical biosensors for cancer diagnosis has been proposed by Gao *et al*.[70] concerning the detection of α-2,6-sialylated glycans, by Zhang *et al*.[71] regarding the determination of β-galactoside α-2,6-sialyltransferase, and by Chen *et al*.[72] for the recognition of Vangl1. Moreover, Xu *et al*.[73] reported its role as anchoring moieties for hollow Pt nanospheres in the fabrication of an amperometric biosensor for the determination of glucose. Finally, Zhang *et al*.[74] used its excellent filming properties for the construction of an electrochemiluminescent biosensor based on graphene quantum dots for the detection of microRNA.

Similar to **PTC-NH$_2$** but obtained from the EDC/NHS coupling of **PTCA** with arginine (**PTC-Arg**) and lysine (**PTC-Lys**), two additional perylene compounds were reported as signal enhancers in graphene-based sensors. Namely, Zhuo *et al*.[75] presented an electrochemiluminescent immunosensor for the identification of α-fetoprotein, a cancer biomarker, containing a **PTC-Arg**/graphene hybrid while Yu *et al*.[76] described an electrochemiluminescent aptasensors based on **PTC-Lys** for the detection of thrombin.

Additionally, Chen *et al*.[77] developed a signal-off electrochemilunescent biosensor for the detection of microRNA based on target induced strand displacement amplification mediated by phi29 polymerase. In their work, **PTCA** was covalently attached to ruthenium(II) tris(2,2'-bipyridyl) by means of tris(2-aminoethyl) amine and served as a bridge to provide the attachment of the metal complex onto GO nano-sheets in order to afford a high initial signal to improve the sensitivity of the biosensor. Based on an analogous protocol, Zhu *et al*.[78] connected a mono-β-cyclodextrin to **PTCA** adsorbed on the surface of rGO to prepare an electrochemical sensor for the detection of persistent organic pollutants in water samples.

## 2.3 Perylene Diimide Based (PDI) Surfactant Systems

As outlined above, apart from **PTCA**-based derivatives, also several **PDI**-based molecules have been designed for the tailor-made functionalization of graphene substrates. For example, Yang *et al*.[79] synthesized a perylene thiol derivative (**ET-PDI**) to improve the dispersibility and stability of rGO in aqueous solution.

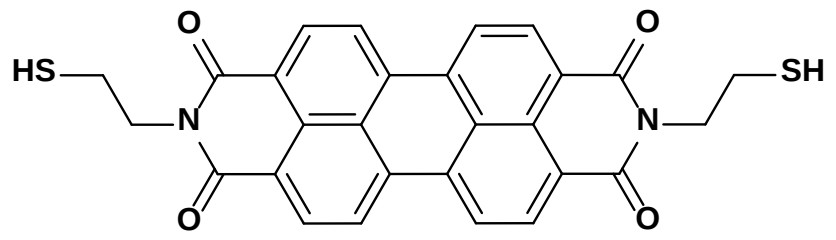

**ET-PDI**

Related to the work of Li *et al*.,[17] the **ET-PDI** was developed to direct the uniform decoration of the rGO sheets with Au-NP by *in situ* nucleation. However, this last step was additionally aided by mercaptoacetic acid in order to passivate the exposed surfaces of the Au-NP and avoid aggregation by cross-linking of the thiol-functionalized rGO sheets. Interestingly, no reducing agent was employed in this step and the growth of Au-NP was triggered only by the rGO substrate.

Yu and coworkers,[80] inspired by the previous work of the group of Loh,[21-22] built up PDI-based wires *via* a non-solvent nucleation protocol and coated them with a shell of rGO by simple solution mixing. Upon analysis of their optoelectronic properties, these novel PDI/rGO nanowires were found to exhibit a peculiar ambipolar charge transport while those obtained without adding rGO exhibited well-defined unipolar *n*-channel performances.

Moreover, Pathipati *et al*.[81] employed a commercially available bay-functionalized PDI (**PDI-CN2**) for the liquid-phase exfoliation of natural graphite in chloroform and investigated the electron mobility of homogeneous **PDI-CN2**/graphene blends compared to thin films of **PDI-CN2** alone. The presence of the exfoliated graphene nano-sheets profoundly influenced the morphology of the **PDI-CN2** deposits as revealed by AFM measurements. The authors suggested that the graphitic flakes served to decrease the number of defect sites in the blend generating a percolation network for a successful charge transport in the disordered regions. Because of these reasons, the **PDI-CN2**/graphene hybrid was characterized by an exploitation of the charge transfer properties compared to **PDI-CN2** alone. In particular, the electron mobility could be increased about 2,000 times as confirmed both by current-voltage and time-of-flight photocurrent experiments.

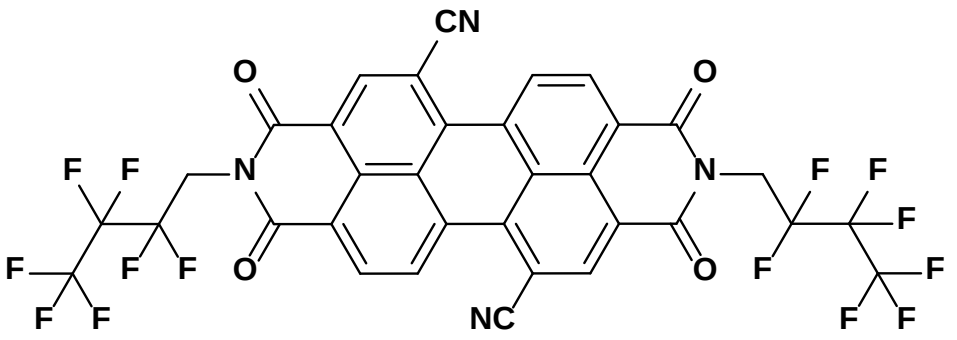

**PDI-CN2**

Furthermore, Bausi *et al*.[82] performed the liquid-phase exfoliation of natural graphite with the aid of an amphiphilic PDI (**amph-PDI**), in order to prepare semi-transparent thin films of graphene. In their experiments graphite was exfoliated in ethanol and a thin film was obtained by *vacuum*-filtration of a centrifuged dispersion followed by transfer on a glass slide. As a comparison, a thin film obtained from the spin-coating of a dispersion of GO was prepared. The AFM characterization revealed the product of the liquid-phase exfoliation to be a thick material (thickness ≈ 10 nm) while the deposited GO material consisted mostly of SLG flakes. However, the sheet resistance was determined to be very high for GO films compared to those obtained with the **amph-PDI**/graphene composite (ca. 21 times higher). Moreover, the conductivity of both films increased after a thermal treatment at 400 °C at very low pressure since this procedure could induce a partial reduction of GO as well as a partial removal of the solvent and of the **amph-PDI** from the surface of the film. Finally, both thin films were modified by addition of trifluoromethanesulfonimide. The results of the chemical doping showed to be limited to the very first layer. Nevertheless, the addition of the doping agent contributed to a strong modulation of the work function in both thin films. For these reasons, the authors envisaged feasible application of these **amph-PDI**/graphene hybrids for the preparation of diodes and optoelectronic devices.

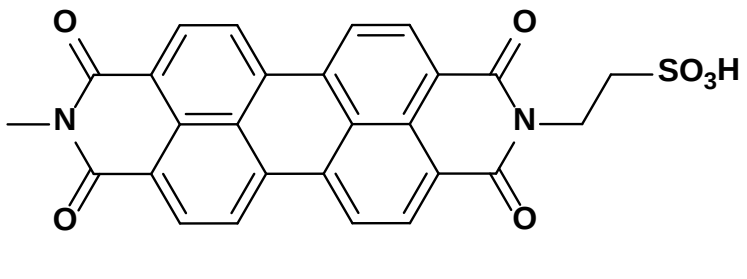

**amph-PDI**

Recently, He *et al*.[83] reported the use of GO as catalyst for the reduction of a bay-functionalized PDI (**SF-PD**) with triethylamine. Unfortunately, the role of GO in the mechanism of this reaction was not explained by the authors. Instead, a detailed photo-physical characterization of the reduced **SF-PDI**/GO hybrid was presented which highlighted an increased solubility of the hybrid in several organic solvents and a remarkable ambient stability.

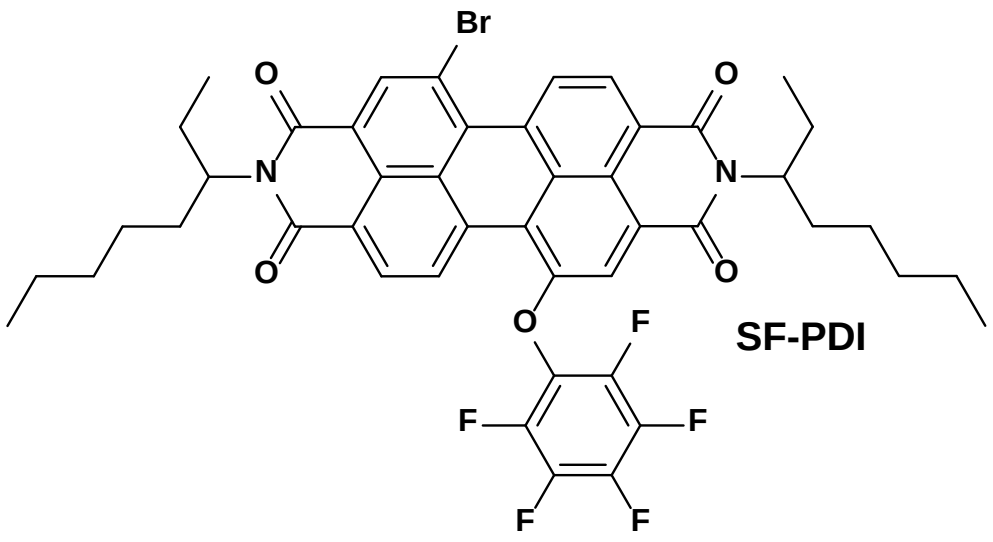

So far, various PDI-derivatives bearing different functional groups have been presented as functionalizing reagents for graphene. Now, the category of PDI-containing polymers will be analyzed. As already summarized for other perylene derivatives, most of the applications involved the incorporation of the macromolecules with GO- or rGO nano-sheets.

In the work of Xu *et al*.[84], a PDI-containing poly(glyceryl acrylate), **PGA-PDI** was allowed to self-assemble onto rGO nanosheets by π–π stacking interactions in order to improve the stability of these dispersions in distilled water. The successful attachment was proven by means of XPS, optical spectroscopy as well as by thermal gravimetric analysis (TGA). Finally, the high biocompatibility and very low citotoxicity of the **PGA-PDI**/rGO composite were evidenced by incubation with 3T3 fibroblasts. Chen *et al*.[85] proposed, instead, the use of a PDI-modified fluorinated poly(hydroxyamide) (**6FPHA-PDI**) for the functionalization of rGO. The **6FPHA-PDI**/rGO complex could be blended with poly(benzobisoxazole) (PBO) to explore and modulate the formation of luminescent materials. Besides a higher thermal stability than PBO alone, the **6FPHA-PDI**/rGO/PBO composite showed also some other interesting photo-physical properties, such as a lower energy gap than PBO, the presence of energy transfers between the PBO and the **6FPHA-PDI** unit as well as a red-shift in the photoluminescence depending on the content of the **6FPHA-PDI**/rGO hybrid employed in the mixture. Moreover, a PDI-containing poly(*N*-isopropylacrylamide) (**PNIPAM-PDI**) was adopted by Wang *et al*.[86] to fabricate a thermo-responsive hybrid with GO and applied for the removal of organic pigments from water solutions. Although the separation process is greatly simplified by using the **PNIPAM-PDI**/GO composite, since the nano-adsorbent could be removed from the solution by simply increasing the temperature above 36 °C, its adsorption capacity is lower than that of pure GO. Nevertheless its removal efficiency is very high (> 99.5 %) a fundamental prerequisite for its future application.

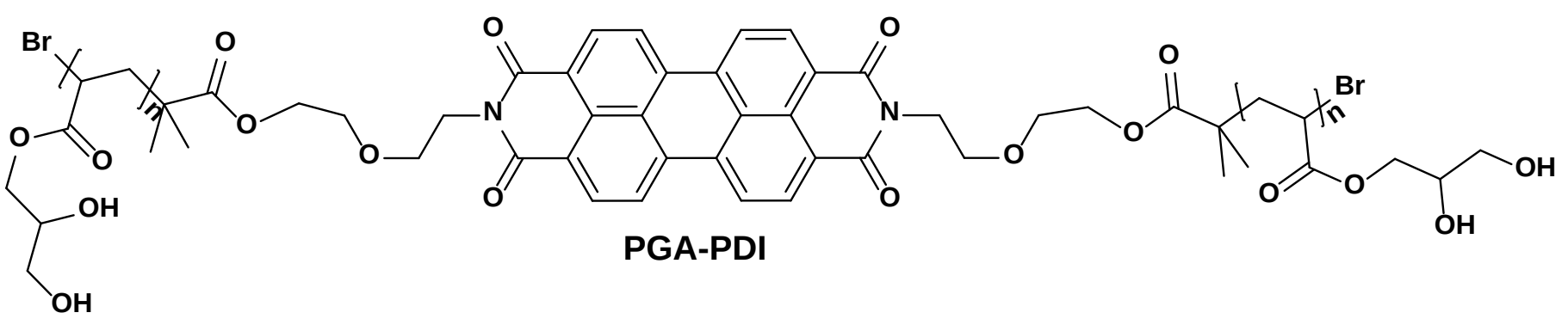

**PGA-PDI**

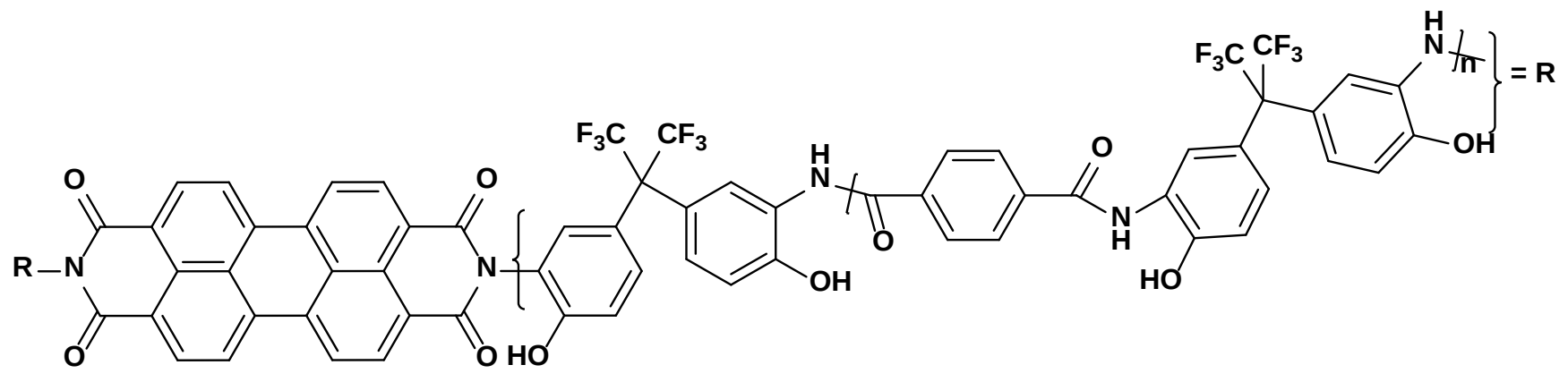

**6FPHA-PDI**

Additionally, Liu *et al*.[87] prepared a perlyene-containing polyglycidol by Steglich esterification of

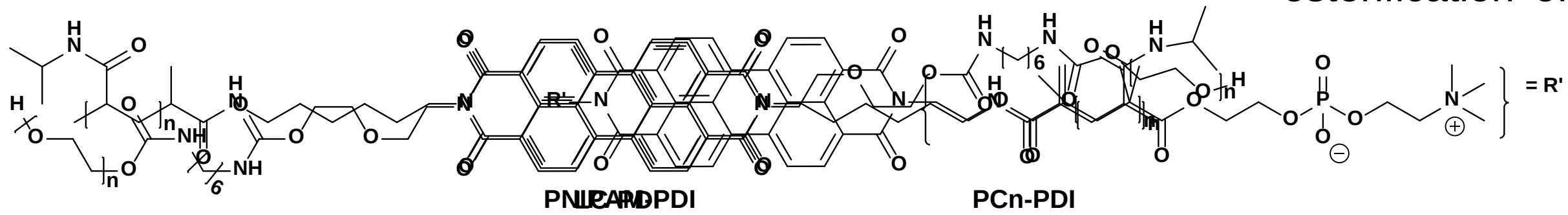

**PCn-PDI**

**PTCA** to enhance the dispersibility of rGO in water and DMF. In these conditions, the dispersions were found to be stable over 2 months. Interestingly, however, the stability of the dispersions in other polar protic and aprotic solvents was comparatively lower (aggregation within 3 weeks). Unfortunately, no further investigations along this line have been provided by the authors. A few years later, Liu *et al*.[88] conceived a stable conjugate of a PDI-containing phosphorycholine oligomer (**PCn-PDI**) with rGO as a vehicle for an anti-tumor agent, paclitaxel (PTX). The **PCn-PDI**/rGO hybrid proved to have a good biocompatibility and the **PCn-PDI**/rGO/PTX composite showed acceptable anti-tumor properties, although with a higher relative proliferation in comparison to PTX alone. These results suggested that the **PCn-PDI**/rGO hybrid could reduce the citoxicity of PTX and the **PCn-PDI**/rGO/PTX was defined as beneficial for application in situations where the reduction of the side effects of PTX might be necessary. In the meanwhile, Zeng and coworkers[89] synthesized a thermotropic liquid-crystal PDI-based polyurethane (**LC-PDI**) and assembled it onto rGO sheets. The resulting **LC-PDI**/rGO nano-structure was then used to improve the thermal and mechanical properties of epoxy resins. A similar liquid-crystal derivative was later proposed by the same group and used in presence of $Al_2O_3$-NP covalently functionalized with (3-aminoprpyl)triethoxysilane.[90] In this latter work the authors could evidence that the presence of both fillers contributed to further enhance the thermal properties of the epoxy resin, such as its conductivity and its glass transition temperature. The authors attributed these additional improvements basically to three factors: i) the strong covalent bonds developed between the $Al_2O_3$ NP and the epoxy matrix during the curing reaction, ii) the uniform dispersion of both fillers, and iii) the bridge-effects afforded by the NP which could fill into the interspace of the entangled PDI/rGO nano-sheets and restrict therefore the mobility of the polymer chains. Finally, He *et al*.[91] created a PDI-containing polyimide for its self-assembly with rGO upon solvothermal reduction of GO in NMP. The authors observed the formation of nano-pin like structures by increasing the content of the macromolecule in the mixture but unfortunately did not provide any explanation for the formation of this interesting phenomenon.

Up to now, only two publications have been focused on the combination of PDI-containing polymers with pristine graphite. On the one hand, in the work of Salavagione *et al*.[92] a polymeric surfactant composed of a hydrophobic PDI dye covalently linked to poly(vinyl alchohol) (**PVA-PDI**) was applied to homogeneously disperse expanded graphite in water. Upon controlled evaporation of the aqueous solution, thin films could

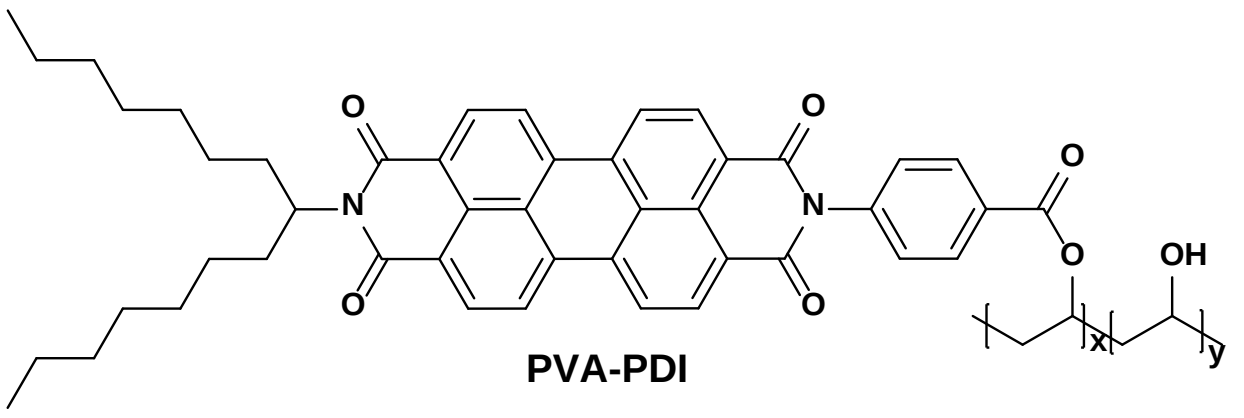

PVA-PDI

be obtained, which were featured for suitable applications in field of flexible electronics. Characterization by SEM showed the presence of a well aligned distribution of few-layer graphene flakes embedded in the polymer matrix. Further investigations by means of differential scanning calorimetry (DSC) showed a complete loss of the crystallinity, a remarkable increase in the glass transition temperature, and an increase of the stiffness of the composite material. The authors suggested that the reason for these dramatic changes are based on the immobilization of the PDI cores on top of the graphene flakes determining a reduction in the segmental mobility of the polymer backbone. On the other hand, in the experiments of Stergiou *et al*.[93] two hybrid structures could be assembled between a photo-active poly-(fluorene-PDI) (**PF-PDI**) moiety with exfoliated graphene sheets by means of both covalent and non-covalent chemistry. The non-covalent functionalization process involved, first of all, two pre-exfoliation steps in chlorosulfonic acid and NMP. **PF-PDI** was subsequently mixed with this pre-treated exfoliated graphitic material and subjected to

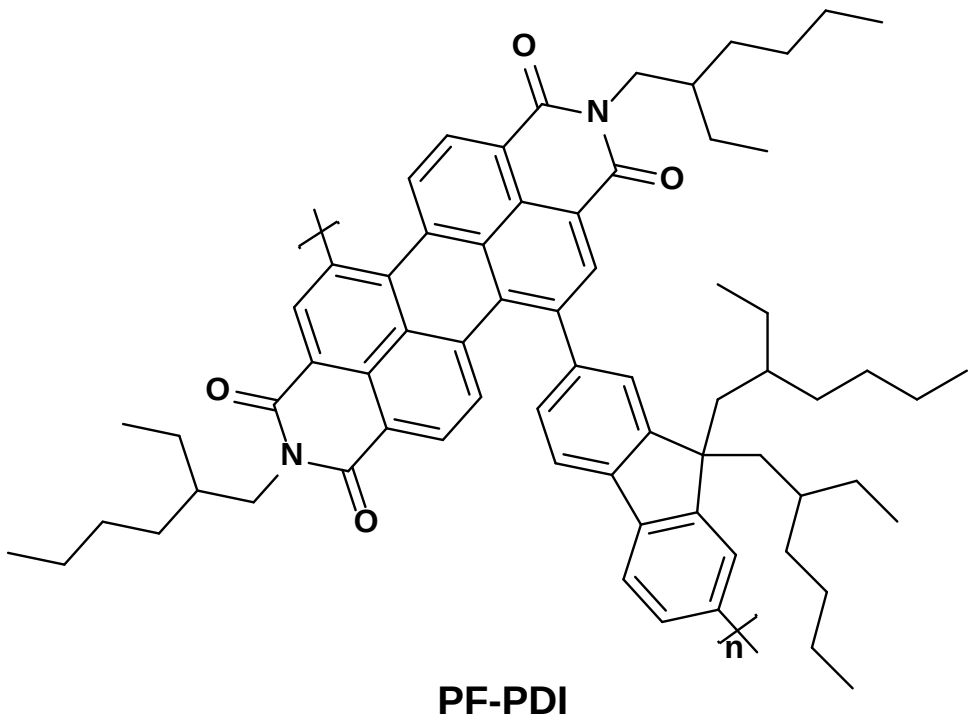

PF-PDI

additional exfoliation in DMF. Intriguingly, the results of the optical and Raman spectroscopic characterization, concerning the **PF-PDI**/graphene hybrid obtained by non-covalent functionalization, showed the presence of a blue-shift of the PDI peaks in the composite and a global n-doping effect, in contrast to the results usually obtained with classic PDIs.[24] Moreover, the covalent hybrid was employed as a catalyst for the reduction of p-nitrophenol in presence of sodium borohydride.

As in the case of **PTCA**, amino acid moieties could be also attached to **PTCDA** and the resultant PDI-based derivatives could be similarly employed for the functionalization of GO and rGO. Roy *et al.*[94] formed a hybrid hydrogel by using tyrosine as amino acid and dissolving the as-prepared PDI derivative (**PDI-C11-Y**) in presence of GO or rGO in buffer solution under sonication. The structure of

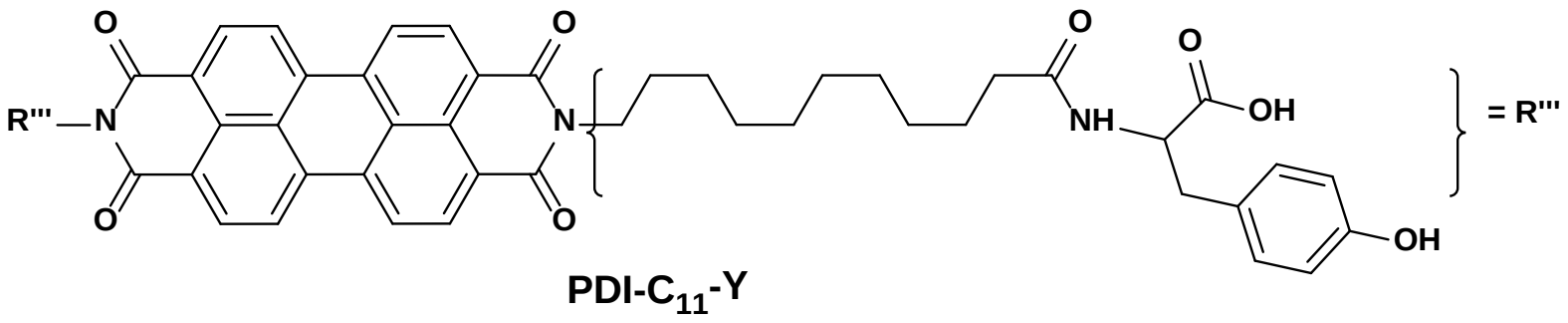

**PDI-C11-Y**

the hydrogel was investigated by means of XRD, which showed the formation of a self-assembled network as result of hydrogen-bonding and π–π stacking interactions, containing well dispersed graphene-based nano-sheets. These results were also confirmed by TEM, which highlighted the coexistence of the cross-linked nanofibrillar network structure of the native hydrogel in the presence of the sheet-like structures of GO and rGO, respectively. Finally, fluorescence and photoconductivity studies implied possible applications for **PDI-C11-Y**/rGO and **PDI-C11-Y**/GO in the field of photo-switching and soft-biomaterials. Additionally, Muthuraj *et al.*[95] attached histidine to **PTCDA** to prepare an assembled structure with GO for the detection of pyrophosphate anions (PPi), as cancer marker. The water-soluble and highly biocompatible **HIS-PDI** probe could efficiently bind copper(II) ions and form oligomeric coordinated structures on the surface of GO, characterized by a complete fluorescence quenching. Upon

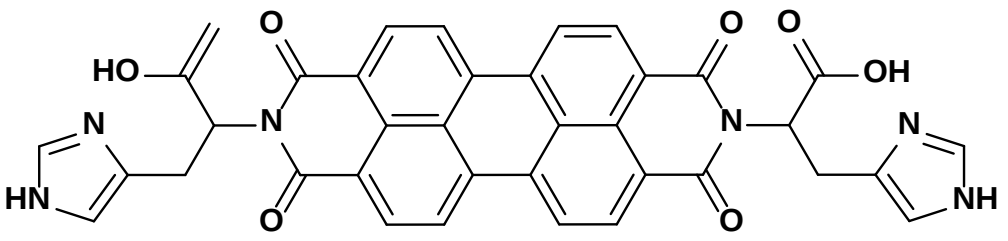

**HIS-PHDIH**

addition of PPi, the metal centers could be liberated and the fluorescence could be restored. Surprisingly, the authors report that concomitantly with the removal of the copper(II) ions by means of the PPi, the **HIS-PDI** was also detached from the surface of the GO nano-sheets. However, the proposed mechanism for this detachment step remains unclear as the provided observations by means of AFM and dynamic light scattering studies

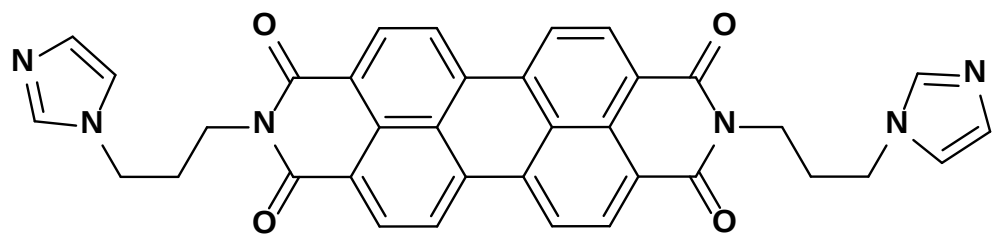

**api-PDI**

remain quite misleading. Balapanuru *et al.*[96] attached, instead, another imidazole-containing molecule, 1-(3-aminopropyl)imidazole, to **PTCDA** in order to build up a new PDI derivative (**api-PDI**) which was able to form a coordination polymer with cobalt(II) ions (**[PDI-Co]n**). This polymer was then immobilized on rGO nano-sheets and the photo-electrochemical activity towards water splitting reactions of the **[PDI-Co]n**/rGO hybrid was studied in detail. These measurements demonstrated that the synergistic interactions between the rGO and the polymeric unit promoted a fast charge transfer and enhanced the yield of hydrogen evolution, thus highlighting the potential of **[PDI-Co]$_n$**/rGO for the use as a photo-catalyst.

By further derivatization of the monomeric form of **api-PDI** with 3-bromopropylamine hydrobromide, Hu and coworkers[97] could synthesize a positively charged bolaamphiphilic PDI molecule (**IL-PDI**). This functional unit was then employed for the decoration of rGO in aqueous solution and led to the formation of a hybrid carrier material which could be used in the fabrication of an electrochemical biosensor for DNA analysis. This device was further improved in the work of Zhang *et al.*[98] where Au-NP were grown *in situ* on the surface of the same **IL-PDI**/rGO hybrid in order to facilitate the electron transfer kinetics of the electrochemical biosensor. Alike, Guo *et al.*[99] developed a new type of PDI-containing ionic liquid (**IL2-PDI**) as anchoring unit for the decoration of graphene sheets with Pd-NP. Due to the high conductivity and electrochemical stability of the **IL2-PDI** moiety, the extremely high specific surface area of the **IL2-PDI**/rGO substrate and the uniform distribution of the Pd-NP, the as-prepared **IL2-PDI**/rGO/Pd-NP composite showed enhanced catalytic activity towards alcohol oxidation and demonstrated interesting applications in direct alcohol fuel cells. Moreover, Li *et al.*[100] referred to **IL2-PD**I for their fabrication of a graphene-based recyclable catalyst decorated with Au-NP, for the hydrogenation of p-nitrophenol in presence of sodium borohydride. In contrast to the study of Stergiou *et al.*[93], the high catalytic activity was attributed to the synergistic effects of both the **IL2-PDI**/graphene carrier and the deposited Au-NP, as the combination of these two elements was argued to dramatically enhance the charge transport and improve the kinetic of the reduction. In addition, Li and coworkers presented also the functionalization of nitrogen-doped rGO (NG) sheets with **IL2-PDI** for the immobilization of Pd-NP and their application as catalyst in direct alcohol fuel cells.[101] Although the authors defined the **IL2-PDI**/NG hybrid as an upgraded substrate for the growth of the Pd-NP, a careful analysis of the electrochemical measurements showed that the electrocatalytic activity of the **IL2-PDI**/NG catalyst is lower compared to that obtained with the **IL2-PDI**/graphene substrate developed by Guo and coworkers.[99]

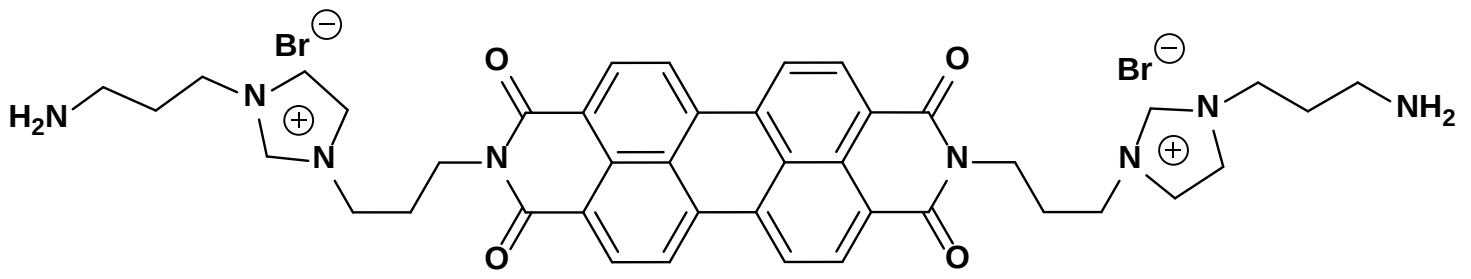

**IL-PDI**

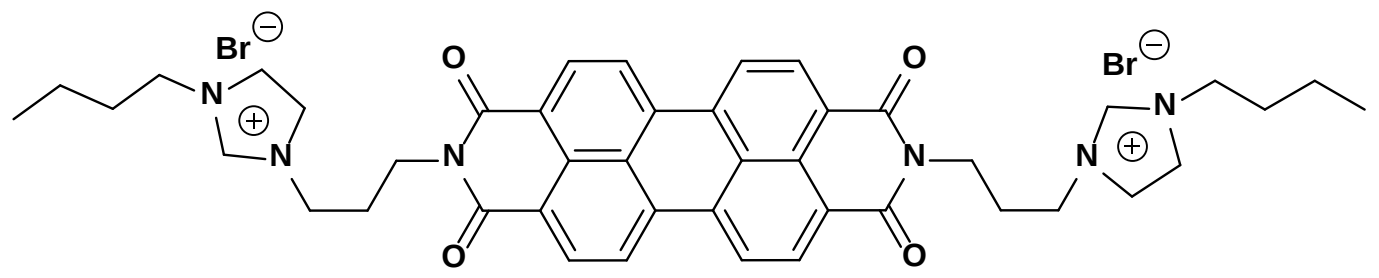

**IL2-PDI**

Recently, Marcia *et al.*[102] employed a Newkome dendronized polycationic PDI (Figure 5) to investigate the mechanism of the non-covalent exfoliation and functionalization of turbostratic graphitic flakes in pure water. As described in detail in their work, the formation of positively charged functionalized 2D graphene-based hybrids could be successfully achieved after carefully optimizing the reaction conditions and the graphite/**PDI** ratio. Furthermore, the outstanding stability against washing of these PDI-functionalized few-layer graphene nano-sheets offered the possibility to incorporate them, together with anionic 0D ZnO-NP building blocks, in highly integrated organic-inorganic architectures by means of a straightforward and versatile iterative dip-coating process. The full characterization of these hybrid supramolecular structures revealed the formation of ultra-thin films with outstanding thickness, centimeter-scale homogeneity, and with an entangled structure exhibiting an intimate contact between the organic and inorganic building blocks. Due to their low cost and high processability, the authors envisaged applications of such hybrid nano-architectures in the fields of energy storage and conversion.

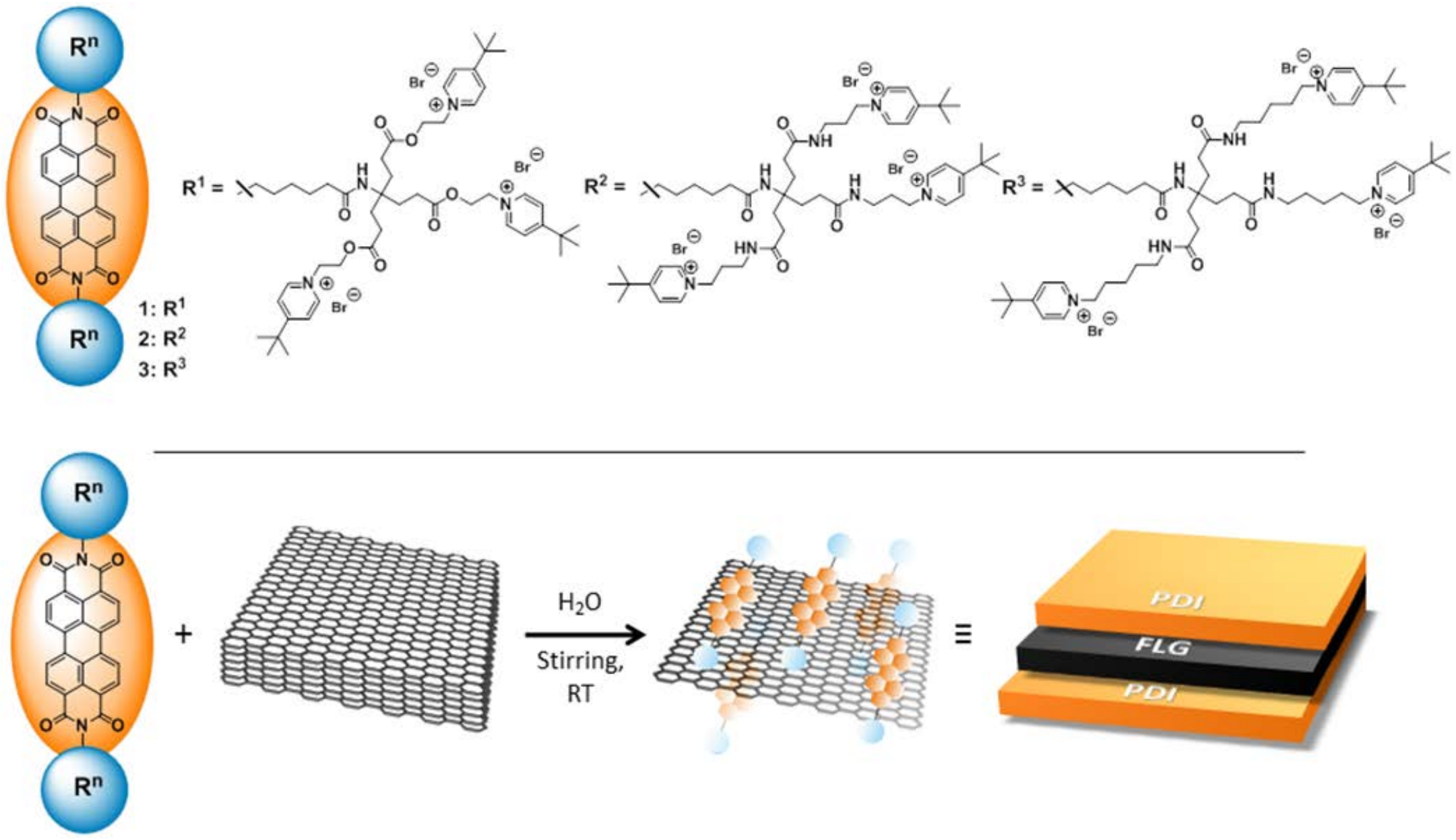

**Figure 5.** Top: Water soluble perylene diimide (PDI) based surfactants for the non-covalent functionalization of graphene. Bottom: Graphical representation for the formation of PDI-functionalized few layer graphene (FLG) by PDI-assisted exfoliation of graphite – no ultrasonic treatment is needed for the graphite delamination process. Ref.[102] with permission from Wiley-VCH Verlag GmbH & Co, Copyright 2016.

Lately, another cationic PDI-based molecule (**TAI-PDI**) was used by Supur *et al.*[103] for its assembly with GO towards the formation of highly ordered supramolecular gel-like structures in aqueous solution. The deposition of **TAI-PDI** on the GO sheets led to a profound modification of the absorption features of the dye

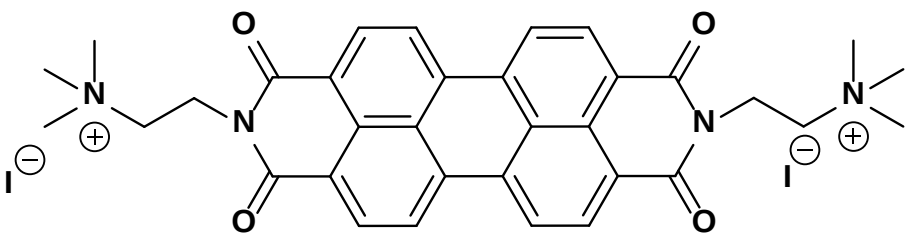

**TAI-PDI**

and a complete fluorescence quenching as a result of the photo-induced electron transfer reactions, which were revealed *via* electrochemistry and femtosecond laser-induced transient absorption spectroscopy. By dispersing **TAI-PDI** and GO in an aqueous solution of polyethylene glycol, the formation of the gel was hindered. This modification of the environmental conditions translated into a dramatic change of the electron transfer processes within the **TAI-PDI**/GO hybrid as a result of its different morphology, which demonstrated that ordered assemblies were responsible for the interlayer charge migration and slow charge separation. Recently, the same authors developed this concept further and proposed a multi-component architecture by means of **TAI-PDI**, a cationic porphyrin, and an anionic Zinc(II) phthalocyanine assembled with GO.[104] This novel 3D-ordered material was designed as broadband light harvester for applications in the field of energy storage and conversion. Because of the residual charges and the respective different molar absorptivities, the authors firstly had to optimize the molar ratio of the three dyes in order to successfully integrate them in an organized self-assembled structure. Moreover, femtosecond laser-induced transient absorption spectroscopy was employed again to monitor the photo-induced electron transfer process within the multi-functional hybrid. Once again a very fast charge separation could be revealed and the authors showed that GO acted only as carrier layer to accelerate the electron transfer between the donor porphyrin and the acceptor (**TAI-PDI**) moiety. This time, however, the charge transport was found to proceed along the lateral dimension whereas it was ineffective in the vertical direction, in contrast to the previous report concerning the assemblies of solely **TAI-PDI** with GO.[103] As outlined above, Newkome dendronized PDI derivatives, like **PDI-2G**, have freequently been used for the dispersion and stabilization of carbon allotropes in aqueous solution. Recently, Berner *et al*.[105] investigated the packing density of a smaller parent derivative of **PDI-2G**, the 1$^{st}$ generation Newkome-dendronized **PDI-1G** for the non-covalent functionalization of single layer graphene grown by chemical vapor deposition (CVD).

After deposition by drop-casting of a solution of **PDI-1G** at very high concentrations, with subsequent washing and drying, several characterization tools were adopted to analyze the structure of the as-adsorbed layer. First of all, the packing density of the molecular layer could be defined, by means of Raman spectroscopy, by a direct comparison of the intensity of the major Raman peaks of **PDI-1G** compared to the G-band of graphene. Moreover, it was also demonstrated that the packing density could be further increased by deposition of **PDI-1G** on pre-annealed graphene substrates. Subsequent analysis *via* STM revealed that **PDI-1G** is adsorbed on the pre-annealed CVD-graphene in a vertical alignment where the PDI cores are somehow tilted with respect to the graphene substrate. This peculiar observation was justified by the extremely high concentration of the dye in solution. Furthermore, fluorescence imaging and -spectroscopy measurements showed the presence of a strongly bathochromic shifted perylene fluorescence, which was attributed to highly aggregated species of **PDI-1G** packed on the graphene substrate. Additionally, Winters *et al.*[106] proposed the direct functionalization of as-grown graphene with **PDI-1G**. Afterwards, this highly non-covalent functionalized material could be easily transferred to other arbitrary substrates without a disruption of the adsorbed layer. In addition, the carboxylic acid groups of **PDI-1G** were used for a subsequent derivatization with ethylene diamine. Raman spectroscopy, XPS, and fluorescence microscopy measurements corroborated the success of this reaction step, which represent the very first example of surface-confined chemistry with a PDI derivative on top of a graphene substrate.

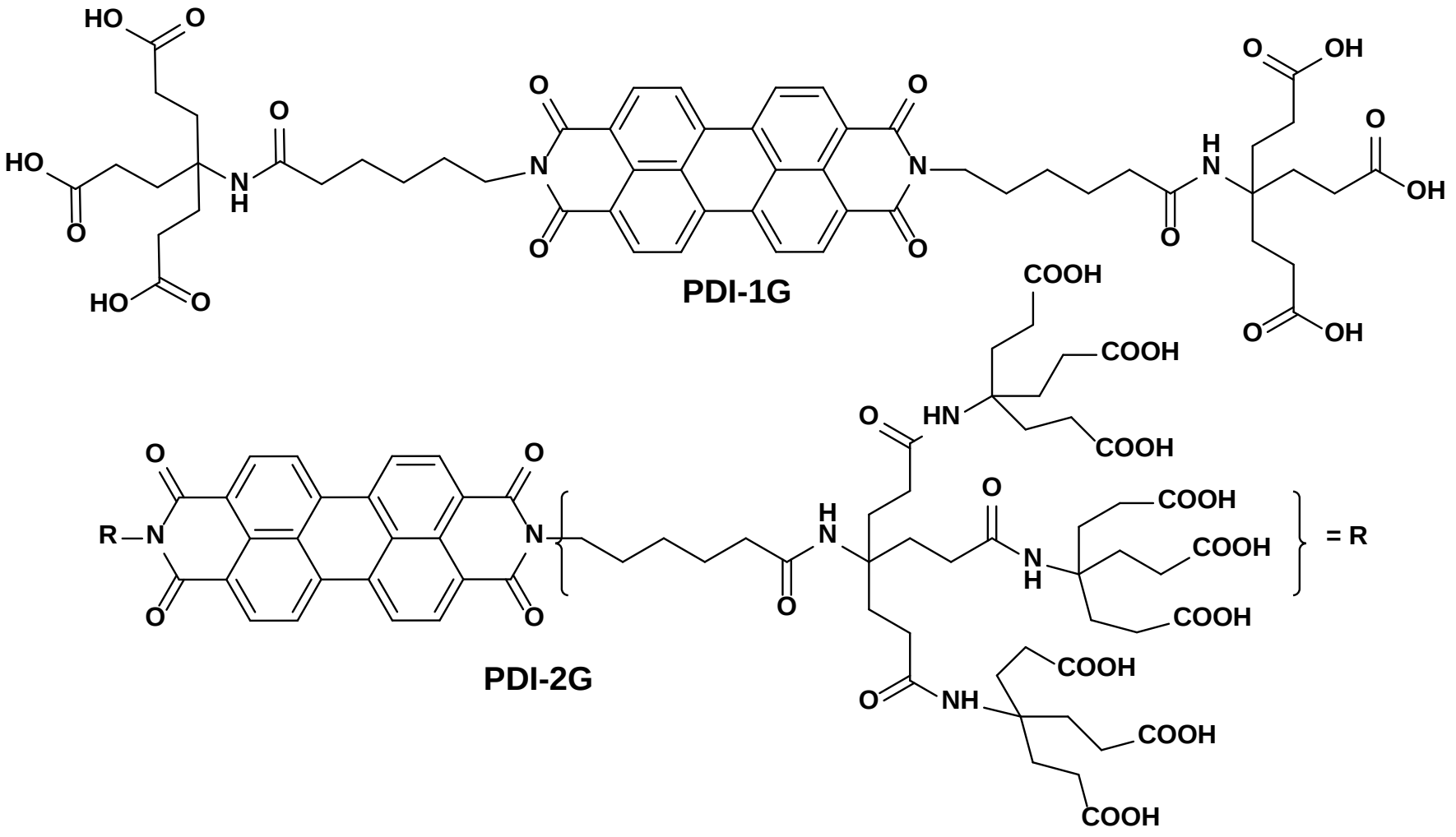

**2.4 Non-Covalent Functionalization of Inorganic 2D Materials**

So far, the integration of both perylene- and PDI-based molecules in the fabrication of graphene hybrid architectures has been described. Triggered by the fundamental research carried out on graphene, within the last years many other inorganic layered crystals have been re-discovered and their exfoliation towards the synthesis of novel 2D nano-materials is being now intensively pursued.[107] Furthermore, the functionalization of these substrates with PAHs is now emerging, as well, as it will play an important role for the establishment of future applications.

The structure of **PTCDA** deposited on $MoS_2$ substrates, has been investigated in the solid state by means of STM techniques since the early 1990s.[108] In addition, AFM and STM images of a bimolecular **PTCDI**-based porous structure were reported recently by Korolkov *et al*.[109] on both hexagonal boron nitride (h-BN) (Figure 6) and molybdenum(IV) disulfide ($MoS_2$). Instead, Zheng *et al*.[110] examined lately the pattern of self-assembled **PTCDA** on top of a single layer of $WSe_2$, which was in turn deposited onto graphite. Furthermore, both perylene- and PDI-based derivatives were employed as seeding layer for the growth of alumina on graphene[111] and on transition metal dichalcogenides[112] *via* atomic layer deposition.

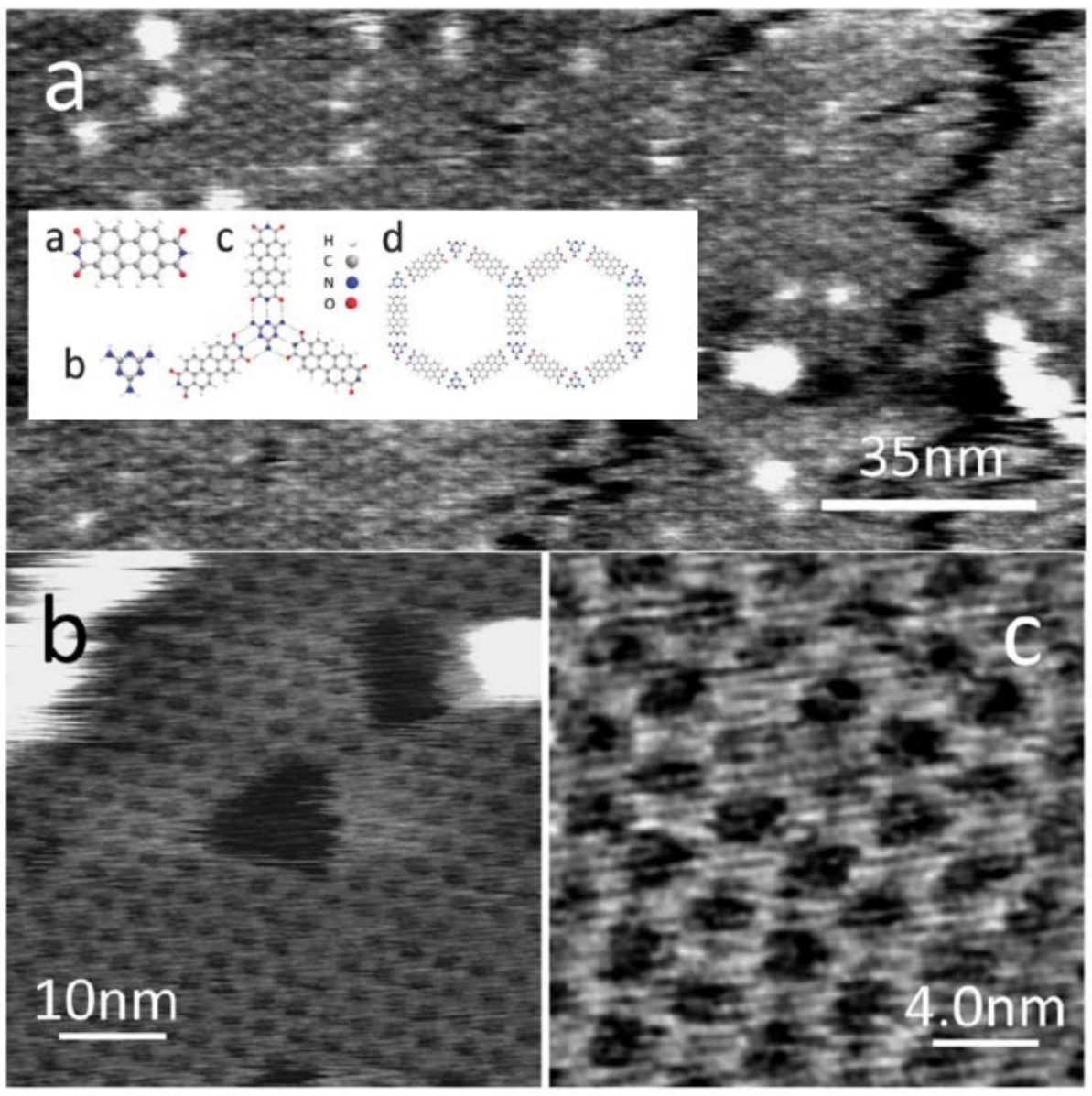

**Figure 6.** AFM scans of PTCDI–melamine networks on the hBN surface in air. Inset: Chemical structure of (a) PTCDI, (b) melamine; (c) schematic of PTCDI–melamine hydrogen bonded vertex, (d) unit cells of the derived supramolecular network. Ref.[109] with permission from The Royal Society of Chemistry, Copyright 2016.

In addition, the potassium salt of **PTCA** was adopted as seeding layer for the growth of transition metal dichalcogenides on various surfaces[113] and also to build up graphene/$MoS_2$[114-115] as well as h-BN/$MoS_2$[116] heterostructures by means of CVD methods. Other highly interesting structures were obtained by Chu *et al*.[117] who was able to nucleate Au-NP on the surface of $MoS_2$ covered with **PTCDA**.

As already explained for graphene-based materials, the synthesis of self-assembled nano-structures *via* physical techniques is generally not straightforward and quite expensive. Therefore wet chemical approaches have to be developed, as well, for the non-covalent functionalization of 2D compounds in solution.[118]

Lately, Marcia *et al*.[119] proposed the synthesis and characterization of a novel class of PDI derivatives, analogues of EDTA (**EDTA-PDI**). Due to their good solubility in water these type of PDI derivatives represent a suitable alternative with respect to the Newkome dendronized systems for the aqueous stabilization and non-covalent functionalization of graphene. Moreover, due to their intriguing structure, which combines the electron deficient central PDI core and a chelating polycarboxylic periphery, these systems provide the basis for the efficient dispersion and functionalization of other 2D inorganic materials, especially transition metal dichalcogenides. As a matter of fact, it is known that specific ions (like $Ni^{2+}$ or $Al^{3+}$) can be efficiently bound to the **EDTA-PDI**, which could be in turn used to create inclusion complexes during the exfoliation process and help to hinder the re-stacking of the dispersed inorganic material.

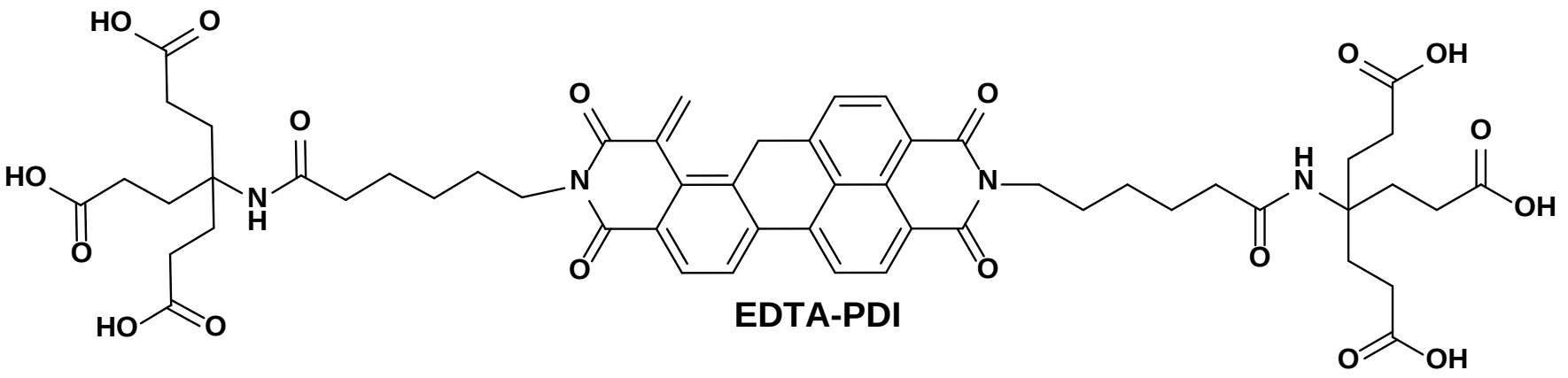

Up to now, only a few reports concerning this non-destructive functionalization of inorganic 2D materials have appeared by using PAH-based derivatives as functionalizing agents. For example a couple of pyrene-based molecules have been employed for the functionalization of h-BN[120-123] and $MoS_2$[121, 124] whereas the use of PDIs has not been reported yet, except of the work of Strauss *et al*.[125] where **PDI-1G** has been adopted as a surfactant for the stabilization of polyhydrogenated graphene in aqueous buffer solution. Perylene-based derivatives, such as the sodium and potassium salt of **PTCA**, have been used by by Zhao *et al*.[126] to exfoliate and functionalize h-BN in aqueous. Moreover, in a recent report of Fu *et al*.,[127] **PTCA** was also conjugated with graphite-like carbon nitride to construct an electrochemiluminescent sensor for dopamine.

Most recently, Abellán *et al*.[128] have reported on the very first approach towards the non-covalent modification of exfoliated black phosphorus by means of PDI-based derivatives in organic solvents. With the aid of a neutral *tert*-butyl protected version of the **EDTA-PDI**, thin phosphorene nano-sheets could be well dispersed and functionalized in THF. The in-depth characterization by means of UV/Vis absorption and Raman spectroscopy as well as electron microscopy (Figure 7) evidenced an increased stability for such hybrid nano-materials which prevents their environmental degradation and thus allows their application for a number of targeting applications.

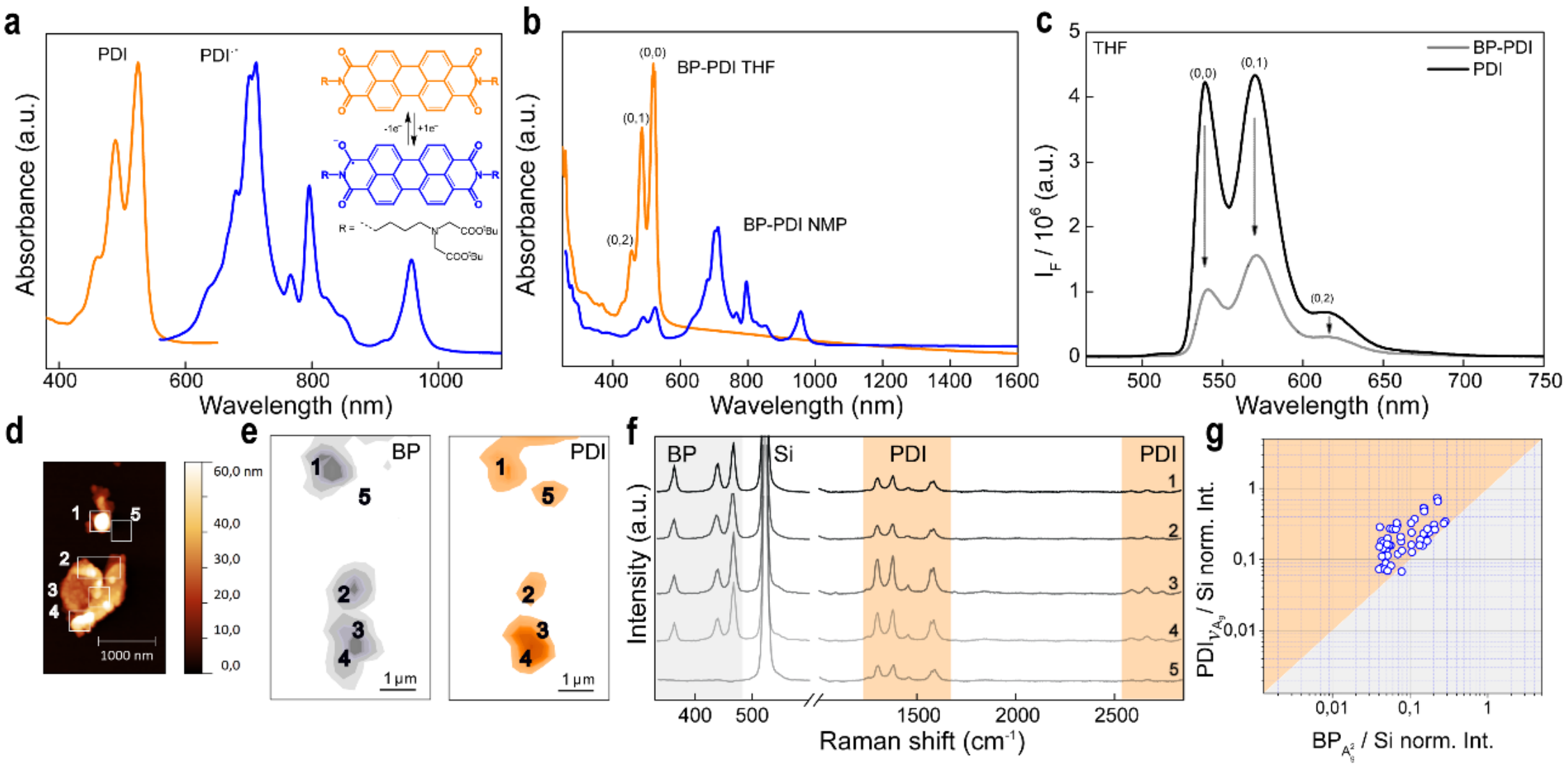

**Figure 7.** a) Molecular structure of the neutral perylene diimide (PDI) and the radical monoanionic one (PDI$^{\cdot -}$). b) Electronic absorption spectra fingerprints of the different PDI species. c) Fluorescence emission spectra of PDI and BP–PDI in THF; $\lambda_{exc}$ = 455 nm. d) Representative AFM image and e) its corresponding Raman $A^1_g$ and f) $\nu_{Ag}$ intensities mappings. (f) Raman spectra of BP–PDI flakes. g) Plot of the $A^1_g$ vs. $\nu_{Ag}$ normalized intensities showing the relationship between the BP and the Raman enhancement effect. Ref.[128] with permission from Wiley-VCH Verlag GmbH & Co, Copyright 2016.

## 3. Summary

In summary, several perylene- and PDI-based derivatives have been successfully integrated in graphene-based materials, which have already found practical applications in different fields of technology. The blooming literature concerning this topic clearly indicates the high versatility of their extended sp$^2$ carbon scaffold as promising building block for the development of non-covalent, supramolecular architectures with graphene and other 2D materials.

Nevertheless, the exfoliation and functionalization of inorganic layered compounds is still at its infancy. Therefore, the ongoing development of novel derivatization concepts with perylene-based surfactant systems is highly demanded. In addition, their availability would directly translate into the possibility of combining several functionalized layered materials into assembled heterostructures in order to fully exploit and tailor their outstanding individual properties.

## Acknowledgement

The authors thank the Deutsche Forschungsgemeinschaft (DFG - SFB 953 "Synthetic Carbon Allotropes", Project A1) and the Interdisciplinary Center for Molecular Materials (ICMM) for financial support. The research leading to these results has received partial funding from the European Union Seventh Framework Programme under grant agreement no. 604391 Graphene Flagship.